\documentclass{emulateapj}
\newcommand{\chan}{\textit{Chandra}}
\newcommand{\xmm}{\textit{XMM$-$Newton}}
\newcommand{\psr}{PSR B1046$-$58}
\newcommand{\eg}{2EG J1049$-$5827}

\begin{document} 

\title{Chandra and XMM-Newton Observations of the Vela-like Pulsar B1046$-$58}

\author{M. E. Gonzalez\altaffilmark{1,2}, V. M. Kaspi\altaffilmark{1,3}, 
 M. J. Pivovaroff\altaffilmark{4}, and B. M. Gaensler\altaffilmark{5,6}}

\altaffiltext{1}{Department of Physics, Rutherford Physics Building, 
McGill University, 3600 University Street, Montreal, QC H3A 2T8, Canada.}
\altaffiltext{2}{gonzalez@physics.mcgill.ca; NSERC PGS B.}
%\altaffiltext{3}{Department of Physics and Center for Space Research, 
%Massachusetts Institute of Technology, Cambridge, MA 02139.}
\altaffiltext{3}{Canada Research Chair, Steacie Fellow.}
\altaffiltext{4}{Lawrence Livermore National Laboratory, Livermore CA 94550}
\altaffiltext{5}{Harvard-Smithsonian Center for Astrophysics, 60 Garden Street,
Cambridge, MA 02138}
\altaffiltext{6}{Alfred P. Sloan Fellow.}

\begin{abstract}
We present results from \chan\ and \xmm\ observations of the radio
pulsar B1046$-$58.
A high-resolution spatial analysis reveals an asymmetric pulsar wind 
nebula (PWN) $\sim$6$''$$\times$11$''$ in size. The combined emission from the 
pulsar and its PWN is faint, with a best-fit power-law photon index of 
$\Gamma$=1.7$^{+0.4}_{-0.2}$ and unabsorbed luminosity of $\sim$10$^{32}$ ergs 
s$^{-1}$ in the 0.5$-$10.0 keV range (assuming a distance of 2.7~kpc). A spatially 
resolved imaging analysis suggests the presence of softer
emission from the pulsar. No pulsations are detected from \psr; assuming a 
worst-case sinusoidal pulse profile, we derive a 3$\sigma$ upper limit 
for the pulsed fraction in the 0.5--10.0 keV range of 53\%. Extended PWN emission is
seen within 2$''$ of the pulsar; the additional structures are highly asymmetric and
extend predominantly to the south-east. We discuss the emission from the PWN as resulting 
from material downstream of the wind termination shock, as outflow from the pulsar or 
as structures confined by a high space velocity. The first two interpretations imply 
equipartition fields in the observed structures of $\gtrsim$ 40$-$100 $\mu$G, while the 
latter case implies a velocity for the pulsar of $\gtrsim$ 190 $n_0^{-1/2}$ km s$^{-1}$ 
(where $n_0$ is the ambient number density in units of cm$^{-3}$). 
No emission from an associated supernova remnant is detected.
 
\end{abstract} 


\section{Introduction} \label{SecIntro}
X-ray observations of rotation-powered pulsars represent a powerful
tool for studying the energetics and emission mechanisms of these objects. 
A large fraction of the available rotational energy is thought to be carried 
away in a relativistic wind of particles. When this wind is confined by the 
surrounding medium it decelerates and a synchrotron nebula forms, called 
a pulsar wind nebula (PWN). 
The overall PWN characteristics provide insights into the content and energy 
spectrum of the pulsar wind, the large-scale magnetic fields, and the 
surrounding medium. The small-scale structures of the nebula provide details 
of the acceleration sites, instabilities in the magnetic field, and the interaction 
between the wind and its surroundings. In young, energetic pulsars, 
nonthermal emission from particles accelerated in the magnetosphere and thermal 
emission from the surface of the star are also expected to be present. 

The radio pulsar (PSR) B1046$-$58 was discovered during a Parkes survey of the 
Galactic plane \citep{jlm+92}. It has a period of $P$ = 124 ms and period derivative
$\dot{P}$ = 9.6$\times$10$^{-14}$ s s$^{-1}$. These values imply a 
characteristic age of $\tau_c$ = $P$/2$\dot{P}$ = 20.4 kyr, spin-down 
luminosity of $\dot{E}$ = 4$\pi^2$$I$$P$/$\dot{P}^3$ = 2.0$\times$10$^{36}$ 
ergs s$^{-1}$ (with a moment of inertia of the neutron star $I$ $\equiv$ 
10$^{45}$ g cm$^{-2}$) and surface dipole magnetic field strength 
of  $B$~= 3.2$\times$10$^{19}$$\sqrt{P\dot{P}}$~G = 3.5$\times$10$^{12}$ G. 
These properties are similar to those of other young neutron stars typified by 
the Vela pulsar. The dispersion measure of 129 pc cm$^{-3}$ towards the 
pulsar implies a distance of 2.7~kpc \citep{cl02}. 

Deep radio observations did not detect extended emission associated 
with \psr\ \citep{sgj99}. X-ray observations with {\it ASCA} and 
{\it ROSAT} detected emission near the pulsar and suggested the presence
of large-scale structures surrounded by faint
emission \citep{pkg00}. However, the poor angular 
resolution of these observations prevented conclusive interpretation of the 
data. \psr\ is also one of a few pulsars with a possible 
EGRET $\gamma$-ray counterpart. The EGRET source \eg\ is coincident with 
the radio coordinates of \psr\ and likely $\gamma$-ray pulsations at the radio 
period have been found \citep{klm+00}.

Here we report on observations of \psr\ carried out with the \chan\ and 
\xmm\ satellites. Our study reveals a faint, arcsecond-scale 
PWN surrounding the pulsar. A detailed 
imaging and spectral analysis of the system suggests the presence 
of soft emission from the pulsar. 
We also examine the characteristics of 
the nebula in light of current models for the production of PWNe.

\section{Observation and Data Analysis} \label{SecObs}
\psr\ was observed with \xmm\ on 2002 August 10. The European Photon 
Imaging Camera (EPIC) MOS and PN instruments \citep{taa+01,sbd+01} 
were operated in full- and small-frame modes, respectively.  These 
settings provide a temporal resolution of 2.6~s for EPIC-MOS and 6~ms for 
EPIC-PN.  The medium filters were used for MOS, the thin filter for 
PN.   The CCD data were reduced with the {\it XMM$-$Newton} Science 
Analysis System (SAS v6.2.0).  The standard screening criteria results in 
an exposure time for the MOS cameras of 20~ks.  For the PN camera, only 
$\sim$70\% of the exposure time was used in the small-window mode 
(4$^{\prime}$ field of view), resulting in an effective exposure time of 15~ks.

The \chan\ observation of \psr\ was carried out on 2003 August 8-9. The aimpoint 
of the back-illuminated ACIS-S3 chip \citep{gbf+03} was positioned at the radio
coordinates of the pulsar. The observation was taken in timed exposure, very faint 
mode, providing a temporal resolution of 3.2~s. The data were reduced 
using the CIAO 3.2.0 software and standard routines. The 
resulting effective exposure time was 36~ks. 

The EPIC-PN instrument
offers a higher sensitivity than ACIS-S, but its lower exposure time and reduced 
field of view limited its use to spectral and temporal analyses of the pulsar
emission. In turn, the \chan\ data were used to perform spectral and high-resolution 
imaging analyses of \psr\ and its surroundings, while the EPIC-MOS observations
were mainly used as a consistency check due to their lower sensitivity and lower 
spatial resolution.

\section{Imaging Analysis} \label{SecImg}
\begin{figure}
\begin{center}
%\epsscale{.80}
%\plotone{f1.eps}
%\includegraphics[width=8.2cm]{}
\includegraphics[width=8.2cm]{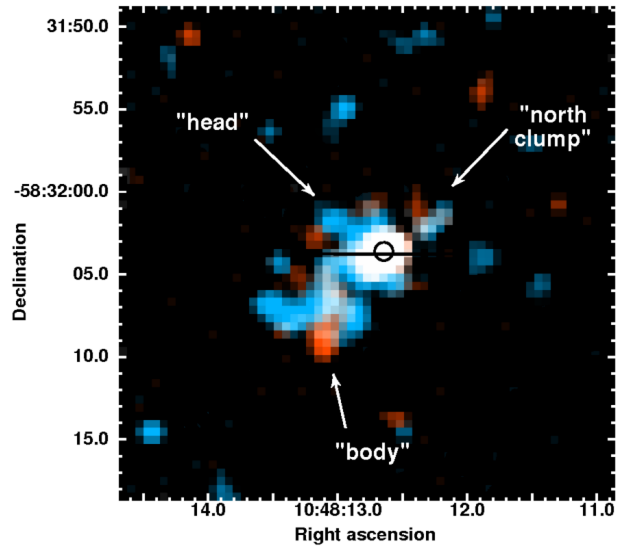}
\caption{\label{figPsrAll} Combined \chan\ image of \psr\ and its PWN using individual
images in the 0.5$-$2.0 keV ($red$) and 2.0$-$10.0 keV ($blue$) ranges. Each 
image was smoothed with a Gaussian with $\sigma$=0$\farcs$5. The circle (0$\farcs$55-radius) and 
line (4$\farcs$7 length) mark
the  X-ray and radio positions of the pulsar, respectively (see \S3). The 
main components discussed in the text are labeled.}
\end{center}
\end{figure}

%To search for extended emission surrounding the pulsar, the high-resolution
%\chan\ data were used.   
Figure \ref{figPsrAll} shows the area surrounding the radio position of \psr\ obtained 
with \chan. The image has been smoothed with a Gaussian with $\sigma$=0$\farcs$5. 
It reveals for the first time the detailed extended structures surrounding \psr, 
which we designate as its pulsar wind nebula (PWN) based on its overall 
characteristics discussed in the next sections. The nebula is elongated and 
$\sim$6$''$$\times$11$''$ in size, with its major axis oriented in a SE-NW direction. 
The line in Figure \ref{figPsrAll} marks the radio coordinates of the pulsar 
at $\alpha_{2000}$=10$^{h}$48$^{m}$12$\fs$6(4$\farcs$7) and 
$\delta_{2000}$=$-$58$\degr$32$'$03$\farcs$75(0$\farcs$02), 2$\sigma$ 
errors \citep{sgj99}. While there
is emission immediately surrounding the pulsar position (PWN ``head''), 
there is also clumped emission to the SE (PWN ``body'') and a bright clump
to the NW (``north clump'').

\begin{figure}
\begin{center}
%\epsscale{.80}
%\plotone{f1.eps}
\includegraphics[width=8.2cm]{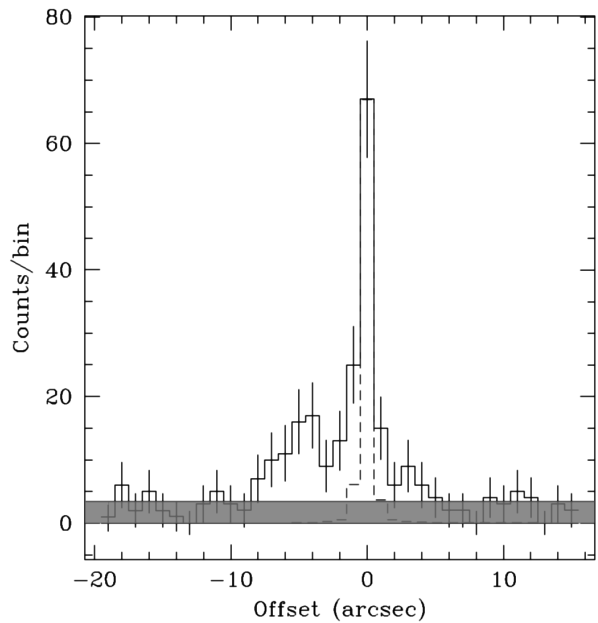}
\caption{\label{figRadPlot} \chan\ axial profile of \psr\ and its PWN in the 
0.5$-$10.0 keV range in a SE-NW direction (solid line) aligned with the implied 
axis of symmetry (see Fig. \ref{figPsrAll}). The origin represents the X-ray position
of the pulsar. The extended emission at offsets of $\sim$$-$3$''$ to $-$8$''$ arises from the
``body" of the PWN (in the SE direction) while the emission at offsets of $\sim$3$''$$-$5$''$ 
arises from the ``north clump". We also show the 2$\sigma$ level of 
the background (shaded region) along with the PSF of the telescope at 1.5~keV 
(dashed line).}
\end{center}
\end{figure}

The head region of the nebula is consistent with a point source surrounded by 
extended structures. Figure \ref{figRadPlot} shows the intensity 
profile of the observed emission along a 10$''$$\times$35$''$ region in
a SE-NW direction aligned with the implied axis of symmetry of the nebula. 
The shaded region represents the 2$\sigma$ background level obtained 
from a nearby, source-free region. The dashed line represents \chan's PSF
at an energy of 1.5 keV obtained using CALDBv3.0.3 and the {\it mkpsf}
tool\footnote{The slight asymmetry in the PSF profile is due to the binning and 
extraction region used in order to match the profile settings for the pulsar. A small,
intrinsic asymmetry in the PSF has been suggested and could also contribute (see 
http://cxc.harvard.edu/cal/docs/cal\_present\_status.html\#psf).}. 
As the head region is also consistent with the radio coordinates of the
pulsar, we consider the source embedded in the region to be the
X-ray counterpart of the pulsar. Comparing the X-ray position of other 
sources in the field with their optical counterparts we derive an X-ray position 
for the pulsar of $\alpha_{2000}$=10$^{h}$48$^{m}$12$\fs$64 and 
$\delta_{2000}$=$-$58$\degr$32$'$03$\farcs$6, with an overall rms error of 
0$\farcs$55. This is coincident within 1$\sigma$ of the radio position, especially
in the more constraining declination direction (see Figure \ref{figPsrAll}). We then 
define the ``pulsar" emission as that arising in a circle of 1$''$ in radius centered 
on the above X-ray coordinates.

\clearpage

We also find that the extended emission from the PWN dominates at high X-ray 
energies. Table \ref{tabCounts}
shows the number of background-subtracted counts obtained for various parts
of the system in the soft (S) and hard (H) bands, their hardness ratios 
(HR = S$-$H / S+H) and their total signal-to-noise ratio \cite[S/N, e.g.,][]{pkc00} in the
0.5$-$10.0 keV range. 
While emission from the nebula is likely to be 
present within the 1$''$ radius used for the pulsar, we suggest that emission from the 
neutron star is indeed present (and may dominate) in this region. The above is supported 
by the fact that this region is the only one that is seen to exhibit a positive HR with high 
significance, suggesting a contribution of distinctly softer emission than that of the rest 
of the nebula.

\begin{figure}
\begin{center}
%\epsscale{.80}
%\plotone{f1.eps}
\includegraphics[width=8.2cm]{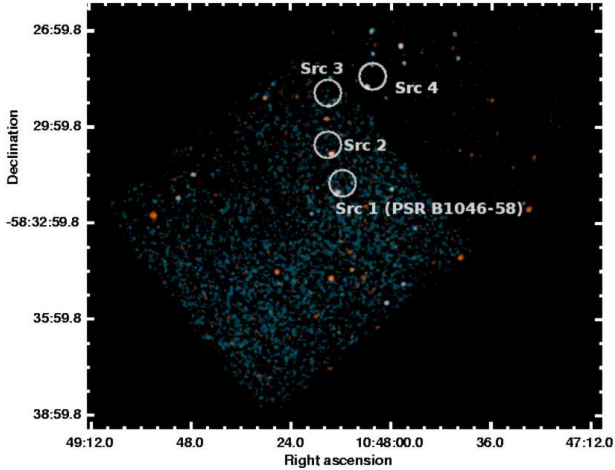}
\caption{\label{figChanField} \chan\ image of the field surrounding \psr\ obtained from the
ACIS-S3 chip and part of ACIS-S2. Individual
background-subtracted and exposure corrected images in the 0.5$-$2.0 keV ($red$) and 2.0-10.0
keV ($blue$) ranges were used. The images were smoothed with a gaussian with 
$\sigma$=2$''$. The circles mark the position of the sources detected with \textit{ASCA} 
and their uncertainties \citep[25$''$ radius, see][]{pkg00}, all of which are coincident with
resolved point sources in the present observations. Similar results (although at lower
angular resolution) were obtained with the \xmm\ MOS instruments. No extended PWN or 
supernova remnant emission was detected in the data.}
\end{center}
\end{figure}

Using \textit{ASCA}, \citet{pkg00} detected four point sources within $\sim$4$'$ of the 
radio position of the pulsar embedded in possible faint emission. 
The \chan\ and \xmm\ observations confirm the presence of these X-ray point 
sources in the field, in addition to many others resolved by \chan\ (see Figure 
\ref{figChanField}). The soft point source in \chan\ coincident with ``Src 2" from 
\citet{pkg00} has an optical counterpart in the USNO B1.0 catalog \citep{mlc+03}, 
while we find no counterparts for the hard sources coincident with ``Src 3" and 
``Src 4" in present catalogues (e.g., USNO B1.0; 2MASS, Cutri et al. 2003\nocite{csv+03}; 
RASS, Voges et al. 2001\nocite{vab+00}). No additional extended emission was detected 
that can be attributed to the PWN beyond the above arc-second structures. 
We then conclude that the low angular resolution and broad PSF wings of the
telescope significantly affected the \textit{ASCA} observations due to the large
number of point sources in the field. We also note that the \textit{ASCA} positions
appear to be systematically north of the \chan\ coordinates, the latter being in
agreement with source positions at other wavelengths.

\section{Spectral Analysis} \label{SecSpec}
The data obtained from both \chan\ and \xmm\ were used to perform a spectral
analysis of the emission from \psr\ and its PWN. For the \chan\ data, an elliptical 
region of 12$''$$\times$18$''$ in size was used with a surrounding annulus as 
background. A total of 184$\pm$14 background-subtracted counts were detected 
in the 0.5$-$10.0 keV range. The \xmm\ data were extracted from circular regions 
of 25$''$ radii in each detector, encompassing $\sim$77\% of the photons from 
a point source, with a nearby region used as background. The data from MOS1
and MOS2 were not included in the analysis due to the low number of counts
detected in them (106$\pm$17 and 47$\pm$17, respectively\footnote{It has been
found that the MOS1 count rate can be artificially boosted by statistical fluctuations
by up to $\sim$70\%, especially for sources with $<$50 true counts. See \S2.1 and 
\S4 in http://xmm.vilspa.esa.es/docs/documents/CAL-TN-0023-2-1.ps}).
The PN detector collected 260$\pm$47 
background-subtracted counts in the same range. The extracted spectra from 
\chan's S3 chip and \xmm's PN detector were fit simultaneously in the 0.5$-$10.0 
keV range using XSPEC (v.11.3.0) and a minimum of 20 counts per bin in each
spectrum. 

\begin{figure}
\begin{center}
%\epsscale{.80}
%\plotone{f1.eps}
\includegraphics[width=8.2cm]{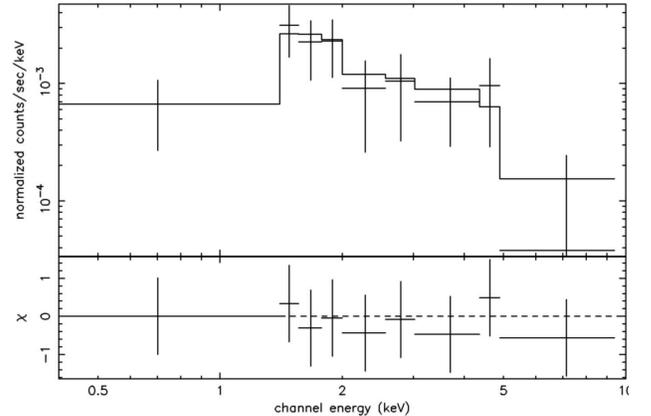}
\caption{\label{figSpec} \chan\ spectrum of the combined emission from \psr\ and its PWN. The
data were binned to contain a minimum of 20 counts bin$^{-1}$. The solid line represents the
best-fit power-law model shown in Table \ref{tabSpecFit} and the lower panel shows the 
fit residuals. }
\end{center}
\end{figure}

Thermal models for the integrated emission from the PWN and \psr\ are
statistically acceptable ($\chi^{2}_{\nu}$$\sim$1.1). However, they 
result in temperatures far too high to represent shock-heated thermal plasma 
\citep[$T$$>$8$\times$10$^{7}$~K, Raymond-Smith model,][]{rs77} or emission 
from the surface of the neutron star ($T$$>$1$\times$10$^{7}$~K, blackbody 
model). Instead, a non-thermal absorbed power-law model 
produces an equally acceptable fit (see Table 
\ref{tabSpecFit}) with a photon index of $\Gamma$=1.7$^{+0.4}_{-0.2}$, similar
to what is observed for other pulsars and their PWNe \citep[e.g.,][]{gs06}. 
The \chan\ spectrum with the best-fit power-law model is shown in Figure 
\ref{figSpec}. The observed emission is thus consistent with having a 
predominantly non-thermal origin.

We then searched for differences in the spectral characteristics of the different
components of the system, as is hinted from our imaging analysis. The high 
spatial resolution available with \chan\ was required in this case. The 116$\pm$14 
counts detected from the PWN alone were fit with a power-law model. Holding
the interstellar absorption fixed to the best-fit value from Table
\ref{tabSpecFit}, the resulting photon index was $\Gamma_{PWN}$ $\sim$
1.2. 
Holding this contribution from the nebula fixed while fitting the overall spectrum, 
the residual (presumably pulsar-dominated) emission was well 
described by a power-law model with a photon index of $\Gamma$ $\sim$ 2.4. 
If a blackbody model is fitted to this emission the resulting temperature is
$\sim$6.1$\times$10$^6$ K, too high to represent purely thermal emission 
from the neutron star but lower than thermal fits to the overall spectrum. 
Although the small number of counts did not allow us to constrain 
the above parameters, emission from the pulsar softer than that of 
the nebula seems to be present.

\section{Timing Analysis} \label{SecTim}
The data from the \xmm\ EPIC-PN camera were used to search for
pulsations from \psr. No evidence for an instrumental 1-s jump in the
data was found \cite[see, e.g.,][]{wkt+04}. The dataset was converted 
to the solar system 
barycenter and a circular region of 25$''$ radius centered at the above 
\chan\ coordinates was used to extract the pulsar events. We also examined 
the dataset in different energy ranges using 0.5$-$10.0 keV,  
0.5$-$2.0 keV and 2.0$-$10.0 keV.
Radio observations of the pulsar 
obtained with the Parkes telescope predict a period for the middle of
our observation (MJD 52496.5) of $\it{f}$ = 8.08512306 Hz 
($\it{P}$ = 123.683955 ms).

The $\it{Z}^{2}_{n}$ test \citep{bbb+83} was used to search for a periodic
signal by folding the extracted photons over a range of 10 trial frequencies 
centered on the radio prediction. The number of harmonics used was varied
to be $n$ = 1, 2, 4, and 8. The most significant signal was found in the
0.5$-$2.0 keV range at $\it{f}$=8.08517(4)~Hz (1$\sigma$ error) with 
$\it{Z}^{2}_{2}$ = 8.9. Although this signal is consistent with the radio
period of \psr, it has a probability of chance occurrence of 6.3\% for a
single trial and it is not significant given the number of searches performed. 
Following 
\cite{vvw+94} and, e.g., \cite{rgs02}, we can derive an upper limit on the pulsed 
fraction. Assuming a worse-case sinusoidal modulation, the maximum Fourier
power obtained for a small range of frequencies centered on the radio
prediction can be used to calculate an upper limit for the pulsed fraction at
a specific level of confidence. In this way, we derive an upper limit for the
pulsed fraction at the 99\% confidence 
level of 53\%, 65\% and 61\% in the 0.5$-$10 keV, 0.5$-$2.0 keV and 
2.0$-$10.0 keV ranges, respectively.

\section{Discussion} \label{SecDisc}

\subsection{\psr}
X-ray emission from young radio pulsars is expected to include contributions from 
any of the following processes: thermal emission from the entire surface due 
to initial cooling, non-thermal emission from magnetospheric processes, and 
possibly thermal emission from localized hot spots reheated by 
back-flowing magnetospheric currents \citep[see, e.g.,][]{krh04}.

The combined emission from the pulsar and its PWN is best described by a 
non-thermal power-law model. However, our imaging and spectral analysis 
suggests that an additional, soft component may be present in the pulsar's 
emission. The soft emission can be well described by a steep power law with
$\Gamma$ $\sim$ 2.4, although such an interpretation would contradict the 
generally observed trend of young pulsars having spectral indices harder than 
those of their PWNe \cite[e.g.,][]{got03}. The
additional soft emission cannot be entirely thermal, as the derived
temperature is too high. It could, however, represent a combined spectrum 
of (hard) non-thermal plus (soft) thermal emission. The number of counts
detected does not allow us to fit two-component models to this emission alone.
Instead, in the case that this overall soft excess is related to thermal 
emission from the whole surface, we 
can constrain its temperature by fitting a blackbody plus power-law 
model and finding the maximum temperature that yields an acceptable fit to the 
total spectrum. We adopt hard limits for the interstellar absorption of 
$N_{H}$ $<$ 2$\times$10$^{22}$ cm$^{-2}$ and a blackbody radius observed
at infinity of $R^{\infty}$ $>$ 10~km (at a distance of 2.7~kpc). The power-law
index and normalization were allowed to vary freely and the maximum blackbody 
temperature that causes deviations from the best-fit model at the 3$\sigma$ 
level was found. This results in a surface temperature of $T^{\infty}_{bb}$ $<$ 
1.4$\times$10$^{6}$~K. This temperature is not constraining for cooling models of
neutron stars, as the temperature in the case of minimal cooling is predicted to 
be $\lesssim$~1.1$\times$10$^{6}$~K for the pulsar's characteristic age 
\citep[e.g.,][]{pgw05}. Similarly, if emission was present from a small polar cap 
on the surface with $R^{\infty}$ $<$ 3~km, the resulting temperature is
$T^{\infty}_{bb}$ $>$ 2.5$\times$10$^{6}$~K. This is comparable to those seen
in other pulsars \citep[e.g.,][]{pzs02} and could then explain the observed soft excess for \psr.

\subsection{The PWN}
The high spatial resolution available with \chan\ allowed us to discover the
arc-second scale PWN structures surrounding the pulsar. The ``head'' of the 
nebula is coincident with the radio position of the 
pulsar, while diffuse emission is seen predominantly to the SE.
The emission from the pulsar and PWN is very faint, with a combined unabsorbed
luminosity of only $\sim$1$\times$10$^{32}$ ergs s$^{-1}$ in the 0.5$-$10.0 keV 
range. The efficiency with which the pulsar converts its rotational kinetic 
energy into X-rays is then $\eta_X$ $\equiv$ $L_X$/$\dot{E}$ $\sim$ 
5$\times$10$^{-5}$, comparable to the values found for other Vela-like systems 
\citep[e.g.,][]{pzs+01,cgg+04}. 

It has been recently suggested that
an empirical relationship exists between the X-ray spectral power-law indices of 
young pulsars ($\Gamma_{psr}$) and their PWN ($\Gamma_{pwn}$) with the pulsar's 
spin-down energy loss $\dot{E}$ \citep{got03}. According to this relationship, lower 
$\dot{E}$ pulsars exhibit harder spectral indeces. 
For \psr, the predicted values are $\Gamma_{psr}$ $\sim$ 0.1 and $\Gamma_{pwn}$ 
$\sim$ 0.9. The PWN emission appears to exhibit a spectrum 
similar to that predicted by this relationship ($\Gamma_{pwn}$ $\sim$ 1.2). However, 
the small number of counts and the possible detection of thermal emission 
limit our ability to constrain the spectral characteristics of the pulsar and its PWN. 
Observations of other Vela-like pulsars do not seem to support this
relationship (e.g., PSR 
B1823$-$13, Gaensler et al. 2003\nocite{gsk+03}). However, longer observations 
at high angular resolution are needed to reliably test the validity of the relationship in this
group of pulsars\footnote{\citet{got04} has later suggested that this relationship only 
holds for pulsars with $\dot{E}$$>$4$\times$10$^{36}$ ergs s$^{-1}$ and without 
a bow-shock morphology.}.

Radio observations did not detect emission from the PWN and constraints on the 
radio properties depend directly on the underlying assumption for the radio efficiency  
\citep[$\eta_R$ = $L_R$/$\dot{E}$;][]{sgj99}. Detected nebulae show efficiencies 
in the range $\eta_R$ $\sim$~10$^{-4}$$-$10$^{-3}$, while upper limits
for unseen nebulae imply values of $\eta_R$ $\lesssim$~10$^{-5}$ \citep{fs97,gsf+00}. 
In the case of an undetected, extended radio nebula surrounding \psr\ and
using a value of $\eta_R$ = 2$\times$10$^{-4}$, the upper limit on the radio flux and 
required surface brightness imply a large radius of 8~pc for a circular nebula \citep{sgj99}. 
At a distance of 2.7~kpc 
this represents a radius of $\sim$10$'$. While the X-ray size of PWNe is often found to be
up to a few times smaller than the radio size due to smaller synchrotron lifetimes in the
X-rays \citep[e.g.,][]{hmb+02},
we find no evidence for extended PWN structures on arcminute scales. For radio sizes 
$\lesssim$~30$''$, based on the small angular size of the X-ray nebula, the required 
efficiencies are very low at $\eta_R$ $\lesssim$~5$\times$10$^{-7}$.

The origin of PWNe is commonly attributed to the interaction of 
a highly relativistic pulsar wind with its surroundings. At a radius $r_s$ from the
pulsar, representing the point of pressure balance
at which the wind is confined and decelerated, we expect $\mathsf{P}$ =
$\dot{E}$/4$\pi$$\Omega$$r_s^2$$c$. Here, $\mathsf{P}$ is the surrounding 
pressure and $\Omega$ $\leq$ 1 is the filling factor of the wind. 
As the head region of the nebula is resolved as an extended structure with \chan, 
we suggest that its outer radius can represent an upper limit on the location of 
the wind termination shock, while the body represents the expected emission 
downstream. The shock radius in this case is $r_s$ $<$ 2$''$ = 0.026 $d_{2.7}$ 
pc, very similar to that found for the Vela pulsar \citep{hgh01} and an order of
magnitude smaller than those of much more energetic pulsars such as the
Crab and PSR B1509$-$58 \citep[e.g.,][]{wht+00,gak+02}. The required pressure 
in this case is $\mathsf{P}$ $\gtrsim$ 8.2$\times$10$^{-10}$ $d_{2.7}^{-2}$ ergs 
cm$^{-3}$ ($d_{2.7}$ is the distance to the pulsar in units of 2.7 kpc). In the region 
downstream of the shock we expect equipartition between the particles and magnetic field 
to be reached, so that $B^2$/4$\pi$ = $\mathsf{P}$, where $B$ is the mean magnetic field 
in the nebula. We then estimate a value of $B$ $\gtrsim$ 100~$d_{2.7}^{-1}$~$\mu$G. The 
corresponding synchrotron lifetime of particles emitting at an energy $\epsilon_{keV}$ 
(in units of keV) is very small, at $t_{\mathrm{synch}}$ $\lesssim$ 
40~$d_{2.7}^{3/2}$~$\epsilon^{-1/2}_{keV}$~yr. The velocity that is needed 
for these particles to reach the edge of the nebula within their synchrotron lifetimes 
is $v$ $\gtrsim$ 2,600 $d_{2.7}^{-1/2}$ $\epsilon$$^{1/2}_{keV}$ km s$^{-1}$ $\gtrsim$ 
0.01~$d_{2.7}^{-1/2}$~$\epsilon$$^{1/2}_{keV}$~$c$. 

One problem with the above interpretation is that we expect the emission downstream 
of the shock to be symmetric about the pulsar. A possible explanation for the lack 
of emission to the NW involves Doppler boosting of the approaching (in this
case SE) component. The observed flux ratios on either side of the pulsar
require an intrinsic expansion velocity of $v\sin\theta$ $\gtrsim$ 0.22$c$, 
where $\theta$ is the inclination of the nebula to the line of sight. Similar velocities
are found in the structures of other PWNe, such as the Crab %and Vela
\citep{hmb+02}.
An intrinsically asymmetric pulsar wind could also be responsible for the observed 
characteristics.

The emission in the body of the nebula could also represent a collimated 
outflow, or jet, along the pulsar spin axis, as observed in other systems such as the 
Crab and Vela \citep{wht+00,hgh01}. In this case, we can estimate the minimum magnetic 
field and energy
needed to support the jet against the surrounding pressure using equipartition arguments 
\citep[e.g.,][]{pac70,shss84}. We assume the jet to be a cylinder with length $l_j$ = 
6.5$''$ = 0.085$d_{2.7}$ pc and radius $r_j$ = 1.5$''$ = 0.020$d_{2.7}$ pc, resulting in an 
emitting volume of $\mathsf{V}$ $\sim$ 3$\times$10$^{51}$$d_{2.7}^{3}$ cm$^{3}$. From the 
results shown in Tables \ref{tabCounts} and \ref{tabSpecFit}, we infer an unabsorbed 
jet luminosity in the 0.5$-$10.0 keV range of $L_j$ $>$ 2$\times$10$^{31}$ ergs s$^{-1}$ 
(1$\sigma$ limit). Equipartition arguments ($B$ $\propto$ $L^{2/7}$$\mathsf{V}^{-2/7}$, 
$E$ $\propto$ $B^{2}$$\mathsf{V}$) then imply  a 
magnetic field in the jet of $B_j$ $>$ 42 (1+$\mu$)$^{2/7}$ $\phi^{-2/7}$ 
$\mu$G and an energy of $E_j$ $>$ 4.4$\times$10$^{41}$ (1+$\mu$)$^{4/7}$ 
$\phi^{3/7}$ ergs. Here, $\mu$ is the ratio of ion to electron energy and $\phi$ is 
the filling factor in the jet. We note that these represent very conservative lower limits on 
$B_j$ and $E_j$, as the total emission from the jet can have a significant contribution at 
lower energies. Using the above limit on $B_j$, the corresponding synchrotron lifetime for 
particles emitting at 2 keV is $t_j$ $<$ 100 (1+$\mu$)$^{-3/7}$ $\phi^{3/7}$ yr and
implies a velocity in the jet of $v_j$ $>$ 1,000~(1+$\mu$)$^{3/7}$~$\phi^{-3/7}$~km~s$^{-1}$. 

For the Crab and Vela, the nebular magnetic field has been estimated to be $\sim$100's 
$\mu$G \citep[e.g.,][]{aabb+04,pksg01}. Similar fields have also been found for the 
large-scale jet in Vela \citep[][]{ptks03}. These are consistent with our derived fields 
above, suggesting the presence of similar confining pressures. The size and energetics 
of the nebula surrounding \psr\ are similar to that of Vela, as expected if these 
properties are also related to the pulsar's spin-down energy $\dot{E}$. In contrast, 
the nebula surrounding PSR B1509$-$58 \citep{gak+02} shows much smaller equipartition
fields ($\sim$8 $\mu$G) and larger size, pointing to a smaller confining pressure.

An alternate argument for an asymmetric nebula involves the confinement of the
pulsar wind due to the ram pressure of a fast-moving pulsar \citep[e.g.,][]{gvc+04}. 
In this case, the body of the nebula would represent the expected X-ray tail aligned 
with the bow shock axis in the direction opposite to the pulsar proper motion. 
The implied position angle in the plane of the sky of the proper motion would be 
$\sim$310$^\circ$ (measured east of north). 
Recent \chan\ observations of the Geminga pulsar revealed a somewhat similar 
morphology for its PWN \citep{psz05,dcm+06} which is likely to arise due the high 
space velocity of the pulsar. 

In the case of a high space velocity, the confining pressure for the nebula is given 
by $\mathsf{P}$ = $\rho$$V^2$, where $\rho$ is the ambient density and $V$ is the 
pulsar's speed. For cosmic abundances and a number density $n_0$ in the ambient
medium we get $\rho$ =  1.37$n_0$$m_H$ = 2.3$\times$10$^{-24}$ $n_0$ g cm$^{-3}$,
where $n_0$ is in units of 1 cm$^{-3}$. Our above estimate for 
$\mathsf{P}$ results in a velocity of $V$ $\gtrsim$~190~$n_0^{-1/2}$~km~s$^{-1}$. 
Such a velocity is consistent with estimates for other bow-shock nebulae 
\citep[e.g.,][]{cc02} and the expected distribution of pulsar velocities at birth 
\citep{antt94,fk05}. The proper motion for the pulsar in this 
case would be $\gtrsim$~0$\farcs$015 $d_{2.7}^{-1}$ yr$^{-1}$ and could be 
detected by radio interferometry. We note that a bow-shock interpretation for the 
X-ray morphology would differ from the static PWN interpretation favoured to 
explain the non-detection of an extended radio nebula \citep{sgj99}. 
A consistent bow-shock interpretation in both radio and X-rays is possible if 
the undetected radio nebula is unresolved with $r_{s,radio}$ $\lesssim$~10$''$ and has 
a low efficiency of $\eta_R$ $\sim$ 10$^{-6}$, which results in estimates from the 
radio observations of 
$V$ $\sim$~250~km~s$^{-1}$ and $n_0$ $\sim$ 0.1 cm$^{-3}$ \citep{sgj99}.

As suggested above, the structures observed in the body of the nebula can represent 
emission due to particles left behind after the passage of the pulsar for a large enough 
velocity. For $V$ $>$ 190 km s$^{-1}$, the time taken by the pulsar to travel the observed 
distance is $<$~540~$d_{2.7}^{2}$ yr. Synchrotron lifetimes matching this limit require 
magnetic fields of $B$ $\gtrsim$ 20 $\mu$G at 1 keV. This is consistent
with the above equipartition field expected to hold in the vicinity of the pulsar given 
the uncertainties in our estimates above.
The emission from the body and the north clump could also contain contributions 
from high velocity
outflows. The north clump at a radius $\sim$ 3$''$ from
the pulsar ``ahead" of the bow is not expected to arise due to shocked ambient medium. 
For example, at a velocity $V$ =
190 $d_{2.7}^{-1}$~km~s$^{-1}$ and density $n_0$ = 1~cm$^{-3}$, the
medium is expected to have a temperature $kT$ $\sim$
0.07$n_0^{-1}$~keV. The 
luminosity is expected to be $L_X$ $\sim$ $\Lambda$$n_0^{2}$$\mathsf{V'}$ ergs~s$^{-1}$ 
$\sim$ 10$^{28}$$n_0^{2}$$d_{2.7}^{3}$~egs~s$^{-1}$, where $\Lambda$ $\sim$ 
1.3$\times$10$^{-24}$$n_0^{-1/2}$ ergs cm$^{3}$ s$^{-1}$ is the cooling function at
$kT$ $\sim$ 0.07$n_0^{-1}$~keV 
and $\mathsf{V'}$ $\sim$ 7$\times$10$^{51}$$d_{2.7}^{3}$ cm$^{3}$ is the
volume of the shocked gas within 3$''$ of the pulsar \citep[see, e.g.,][]{rl79, gvc+04}.
The implied absorbed flux at Earth is then much smaller than the measured values. 
Additional observations can be used to 
confirm the presence of the north clump and its association to the system.

\subsection{\psr\ and \eg} 
Although not confirmed, an association between \psr\ and \eg\ has 
been proposed in light of, among other things, their positional coincidence, possible 
$\gamma$-ray pulsations, and spectral and energetic properties of the $\gamma$-ray 
pulsar \citep{fie95,pkg00,klm+00}. \eg\ was found to have a 
pulsed $\gamma$-ray flux above 400 MeV of (2.5$\pm$0.6)$\times$10$^{-10}$ ergs 
cm$^{-2}$ s$^{-1}$, implying a $\gamma$-ray efficiency of (9$\pm$2)$\times$10$^{-3}$ 
\citep[for 1~sr beaming and a distance of 2.7 kpc,][]{klm+00}. 
Unresolved, unpulsed $\gamma$-ray emission at the pulsar position was also
found. 

An extrapolation of the observed X-ray spectrum for \psr\ and its PWN (see 
Table \ref{tabSpecFit}) to energies $>$400 MeV predicts a flux of 
0.7$\times$10$^{-13}$ ergs cm$^{-2}$ s$^{-1}$, three orders of magnitude lower
than that derived for the EGRET source. This is true for other well-established 
$\gamma$-ray pulsars \citep[e.g.,][]{tbb+99}. Observations at higher X-ray energies 
(e.g., deep exposures using {\it RXTE}, {\it Suzaku}) can help determine whether \psr\ 
is indeed a $\gamma$-ray pulsar by detecting possible high X-ray energy pulsations 
and spectral characteristics consistent with those of \eg.

We note the presence of a particularly hard X-ray source visible in the 
\chan\ data, $\sim$3$\farcm$3 from the pulsar position at coordinates
$\alpha_{2000}$=10$^{h}$48$^{m}$05$\fs$74($\pm$0$\fs$15) and 
$\delta_{2000}$=$-$58$\degr$28$'$46$''$($\pm$2$\farcs$4). The source is
coincident with ``Src 4'' in \citet[][]{pkg00}; see also Figure \ref{figChanField}. 
A total of 164$\pm$15
counts were detected from this source and its spectrum is well described by a
non-thermal power-law model with $N_H$ = (0.7$^{+0.4}_{-0.2}$)$\times$10$^{22}$
cm$^{-2}$, $\Gamma$ = 1.1$^{+0.3}_{-0.2}$ and unabsorbed flux in the
0.5$-$10.0 keV of (1.2$^{+0.5}_{-0.4}$)$\times$10$^{-13}$ ergs s$^{-1}$ cm$^{-2}$
($\chi^2$ = 0.45 for 5 d.o.f., 1$\sigma$ errors). No counterpart for this source was 
found at other wavelengths and it lies within the 
99\% confidence error of \eg. This suggests that the 
observed $\gamma$-ray 
emission could have contributions from more than one source. Higher spatial
resolution $\gamma$-ray observations with future missions may help to address 
this possibility.

We also find that the emission from the nebula dominates at high X-ray energies 
(see Table \ref{tabCounts}) over that of the pulsar. This argues 
for the unpulsed $\gamma$-ray 
emission associated with \eg\ to also arise, at least in part, from the nebula, 
as proposed for the Vela and Crab pulsars \citep{nab+93,kab+94,fmnt98}.

\subsection{Non-detection of a Supernova Remnant}
As we do not detect emission from a supernova remnant (SNR) in the field containing 
\psr, following \citet{gsk+03} we can estimate the flux from a possible unseen remnant 
by scaling the emission observed from the Vela SNR. The emission from this remnant 
can be described by a two-component Raymond-Smith plasma with temperatures 
$kT_1$ $\sim$ 0.15 keV and $kT_2$ $\sim$ 1 keV \citep{la00}. The total unabsorbed 
luminosity in the 0.1-2.4 keV range is 2.2$\times$10$^{35}$ 
ergs s$^{-1}$ and the cooler component has a volume-integrated emission measure 
$\sim$10 times higher than the hot component. Scaling to a distance
of 2.7 kpc, the unabsorbed flux densities for each component would be
$f_1$ = 2.3$\times$10$^{-10}$ ergs s$^{-1}$ cm$^{-2}$ and $f_2$ =
2.2$\times$10$^{-11}$ ergs s$^{-1}$ cm$^{-2}$, respectively.
 
The expected ACIS-S\footnote{The EPIC-MOS data provide less constraining limits 
for this emission due to their lower sensitivity and higher background levels. The 
EPIC-PN data had a much smaller exposure time and limited field of view.} 
count rate predicted by the WebPIMMS 
tool\footnote{http://heasarc.gsfc.nasa.gov/Tools/w3pimms.html} is
$<$2.6 counts s$^{-1}$ (on-axis) for absorbing columns $>$5$\times$10$^{21}$ 
cm$^{-2}$. Scaling the size of the Vela SNR \citep[$\sim$8$^\circ$ at 300 pc,][]{la00}, 
we would expect a shell with a radius of $\sim$26$'$ at 2.7 kpc. This size is 
larger than the ACIS-S (and EPIC-MOS) field of view. However, parts of the 
remnant might be visible in the present data given the uncertainties in the
spatial distribution of the remnant and direction of motion of the pulsar. 
Assuming emission areas for the remnant proportional to having a shell 
thickness $>$20\% of its radius, the expected surface brightness density is 
$\lesssim$0.8$\times$10$^{-6}$ counts s$^{-1}$ arcsec$^{-2}$. This is 
lower than the ACIS-S background level during the observation,
estimated to be $\sim$2$-$6$\times$10$^{-6}$ counts~s$^{-1}$ 
arcsec$^{-2}$ (accounting for vignetting at large
off-axis angles). 

It is then possible that a remnant with similar properties 
to that of the Vela remnant might be present and it is undetected by these
observations.
The lack of radio or X-ray emission from the
SNR has also been attributed to an initial fast expansion into a low-density
cavity \citep[e.g.,][]{bgl89}. This expansion is followed by a collision with a dense 
surrounding shell of stellar wind material from the progenitor  where the emission 
fades rapidly and energy is dissipated through radiative shocks.

\section{Conclusions}
We have detected an asymmetric, arc-second scale PWN surrounding the young,
energetic \psr. 
The overall emission from the 
pulsar and PWN is best described by a non-thermal power-law model with 
$\Gamma$=1.7$^{+0.4}_{-0.2}$ and a low X-ray 
luminosity of $\sim$10$^{32}$ ergs s$^{-1}$ in the 0.5$-$10.0 keV range. 
The brightest emission region in the PWN is coincident with 
the radio coordinates of the pulsar. It is resolved as an extended structure with 
\chan\ and we suggest that it can represent an upper limit to the location of the 
wind termination shock. The additional, 
asymmetric structures can represent nebular emission downstream of the shock or 
collimated outflow from the pulsar. The implied equipartition magnetic fields are
$\gtrsim$~40$-$100 $\mu$G. The overall size and energetics of the system in this 
case are very similar to those of the Vela pulsar. Instead, if the observed 
asymmetry in the nebula were to arise due to a large space velocity for the pulsar we
estimate a value of $\gtrsim$ 190~$n_0^{-1/2}$ km~s$^{-1}$. In this case, we would favor 
the presence of arcsecond-scale nebulae with low efficiencies in both X-ray and radio. 
Our spatially resolved analysis also hints at the presence of softer
emission from the pulsar than from the rest of the nebula. In the case that
thermal emission from the whole surface of the star is present, we derive an 
upper limit for the temperature of $<$1.4$\times$10$^{6}$~K, which is not constraining
on cooling models of neutron stars. In the case of a small hot spot on the surface,
the implied temperature is $>$2.5$\times$10$^{6}$~K and would be consistent with 
those seen for other pulsars.

The results shown here demonstrate the need for high-resolution, high-sensitivity observations 
in order study the wide range of structures associated with rotation-powered pulsars.
While the origin of  these structures cannot be unambiguously determined in the case of
\psr, they can be broadly understood using current theories for the production PWNe.
Additional data and theoretical work are needed to reach a consistent picture 
of this interesting phenomenon in neutron star physics.

\acknowledgments
This work was supported in part by an NSERC Discovery Grant and Steacie Supplement, 
FQRNT, NSERC Graduate Scholarship and a Canadian Institute for Advanced Research 
Fellowship, the Canada Research Chair Program, SAO grant GO3-4068X awarded by the 
CXC and by NASA grant NAG5-11376 awarded by NASA's XMM Guest Observatory Facility. 
B.M.G. acknowledges the support of NASA through LTSA grant NAG5-13032, and of an 
Alfred P. Sloan Research Fellowship. We thank R. N. Manchester for kindly providing the 
Parkes radio ephemeris for \psr.

\bibliography{journals1,psrrefs,modrefs}

\begin{deluxetable}{lcccc}
%\begin{center}
\tabletypesize{\footnotesize}
\tablewidth{0pt} 
\tablecaption{\label{tabCounts} Spatial analysis of \psr\ and its PWN with \chan}
\tablehead{ \colhead{} &\colhead{0.5$-$2.0 keV} &\colhead{2.0$-$10.0 keV} &\colhead{HR = } & \colhead{S/N} \\
\colhead{Region} &\colhead{counts (S)} &\colhead{counts (H)} &\colhead{(S$-$H)/(S+H)} &\colhead{(0.5$-$10.0 keV)} }
\startdata
Pulsar (1$''$ radius\tablenotemark{a})  & 43 $\pm$ 8 & 24 $\pm$ 6 & 0.27 $\pm$ 0.10 & 8.5$\sigma$ \\
PWN, all (- Pulsar)  & 46 $\pm$ 9 & 70 $\pm$ 11 & $-$0.21 $\pm$ 0.06 & 9.1$\sigma$ \\
PWN, ``head'' (- Pulsar) & 12 $\pm$ 5 & 31 $\pm$ 7 & $-$0.44 $\pm$ 0.21 & 6.1$\sigma$ \\
PWN, ``body'' & 22 $\pm$ 6 & 34 $\pm$ 7 & $-$0.22 $\pm$ 0.10 & 9.1$\sigma$ \\
``North clump'' & 5 $\pm$ 2 & 5 $\pm$ 2 & 0.00 $\pm$ 0.02 & 2.9$\sigma$ \\
Pulsar + PWN & 89 $\pm$ 11 & 95 $\pm$ 12 & $-$0.03 $\pm$ 0.01 & 12.1$\sigma$ \\
\enddata
\tablenotetext{a}{See \S\ref{SecImg} for detailed discussion.}
%\end{center}
\end{deluxetable}

\begin{deluxetable}{lc}
%\begin{center}
\tabletypesize{\footnotesize}
\tablewidth{0pt} 
\tablecaption{\label{tabSpecFit} Power-law Fit to combined emission from \psr\ and 
its PWN\tablenotemark{a}} 
\tablehead{ \colhead{Parameter} &\colhead{Value ($\pm$1$\sigma$)}}
\startdata
$N_{H}$ (10$^{22}$~cm$^{-2}$) & 0.9$^{+0.4}_{-0.2}$\\
$\Gamma$ & 1.7$^{+0.4}_{-0.2}$ \\
$\chi^2$(dof) & 79(68)\\
$f_{abs}$ \tablenotemark{b} (ergs~s$^{-1}$~cm$^{-2}$) & 0.7$^{+0.6}_{-0.1}$$\times$10$^{-13}$\\
$f_{unabs}$ \tablenotemark{b} (ergs~s$^{-1}$~cm$^{-2}$)  & 1.0$^{+0.7}_{-0.2}$$\times$10$^{-13}$\\
$L_{X}$ \tablenotemark{c} (ergs~s$^{-1}$) & 0.9$^{+0.6}_{-0.2}$$\times$10$^{32}$\\
\enddata
\tablenotetext{a}{Simultaneous fit to both \chan\ and \xmm\ spectra.}
\tablenotetext{b}{Absorbed and unabsorbed X-ray fluxes, $f_{abs}$ and $f_{unabs}$, in the 
0.5$-$10.0 keV range.}
\tablenotetext{c}{Unabsorbed X-ray luminosity in the 0.5$-$10.0 keV range for a distance
of 2.7~kpc.}
%\end{center}
\end{deluxetable}

\end{document}